\documentstyle[12pt,epsfig]{article}
\title{Rigidity dependent knee and cosmic ray induced high energy 
neutrino fluxes}
\author{Juli\'an Candia$^a$ and Esteban Roulet$^b$\\
$^a${\it Departamento de F\'{\i}sica, Universidad Nacional de La Plata, 
CC67,}\\{\it La Plata 1900, Argentina}\\
$^b$ {\it CONICET, Centro At\'omico Bariloche, Av. Bustillo 9500,}\\
{\it Bariloche 8400, Argentina}}
\begin{document}
\maketitle
\begin{abstract}
Scenarios in which the knee of the cosmic ray spectrum depends on the 
particle rigidities usually predict that the cosmic ray composition becomes
heavier above the knee and have associated a change in the spectral 
slope of each individual nuclear component which is steeper than the 
change ($\Delta\alpha\simeq 0.3$) observed in the total spectrum. We 
show that this implies that the very high energy ($E_\nu>10^{14}$~eV) 
diffuse neutrino fluxes produced by cosmic rays hitting the atmosphere 
or colliding with the interstellar medium in the Galaxy will be 
significantly suppressed, making their detection harder but also 
reducing the background for the search of other (more challenging) 
astrophysical neutrino sources.
\end{abstract}
\section{Introduction}
The study of atmospheric neutrinos of sub-GeV and multi-GeV energies has been 
of paramount importance in the recent past, providing in particular the first 
clear evidence in favor of neutrino oscillations, and hence of non-vanishing 
neutrino masses. These neutrinos are produced mainly in the decay of pions 
and kaons (e.g. $\pi^-\to\mu^-\overline{\nu}_\mu$, with $\mu^-\to e^-
\overline{\nu}_e\nu_\mu$) produced by the cosmic rays (CRs) hitting the upper 
atmosphere and generating air showers with a significant hadronic component. 
Due to the interplay between the mesons' decay and their interactions in the 
air, as the energy increases (with the associated relativistic dilation of 
their decay times) the mesons lose an increasing amount of  energy 
before they decay, and also the muons produced can reach the ground before 
decaying. As a consequence of this, the predicted atmospheric neutrino 
spectrum results steeper than the CR spectrum typically by an extra power of 
the energy, with d$N_\nu/{\rm d}E_\nu\propto E_\nu^{-3.7}$, while d$N_{CR}
/{\rm d}E_{CR}\propto E_{CR}^{-2.7}$ \cite{vo80,ga90,li93}.

Comparing the charged mesons' decay length, $L\equiv\gamma c\tau$, where
in particular $L_\pi\simeq 5.6$~km($E_\pi/100$~GeV) and 
$L_K\simeq 7.5$~km($E_K/{\rm TeV}$), it is clear that above 100~GeV the 
pions traverse several attenuation lengths (corresponding 
to approximately 120~g/cm$^2$) in the atmosphere before decaying, and due to 
this the neutrino fluxes above a few hundred GeVs actually arise mainly from $K$ decays. 
Kaons in turn are also significantly attenuated before decaying for energies 
above the TeV, with the effect that the prompt neutrinos from charmed particle decays become 
increasingly important.

Several groups have studied the charm particle production by CRs 
\cite{th96,pa99,ma03}, which actually
requires to take into account next to leading order (NLO) processes, which 
increase the charm production by more than a factor of two, 
and the cross sections
are sensitive to the values of the partonic distribution functions at very 
small values of the scaling variable $x$, what introduces further 
uncertainties in the results, due to the need to extrapolate beyond
the measured range.
Even if the charmed mesons ($D^\pm,\ D^0,\ D_s,\ \Lambda_c, ...$) 
are produced at a rate $\sim 10^{-4}$ with respect to the non-charmed ones, 
their very short decay time, with $L_D\simeq 2$~km($E_D/10$~PeV), implies 
that the prompt neutrino spectral slope just follows the CR slope, and hence 
the prompt neutrinos eventually dominate the atmospheric neutrino fluxes
above a PeV in the vertical direction, and somewhat above that energy at 
large zenith angles\footnote{At large zenith angles the mesons produced high 
in the atmosphere are less affected by the attenuation before decay due to 
the fact that the atmosphere they traverse is more tenuous.}.

It is important at this point to take into account the fact that the primary 
CR spectrum shows a steepening at the so-called {\it knee}, corresponding to 
$E_{knee}\simeq 3\times 10^{15}$~eV, with d$N_{CR}/{\rm d}E_{CR}
\propto E_{CR}^{-3}$ for $E_{CR}>E_{knee}$. The associated steepening 
in the induced neutrino fluxes has been obtained in the literature 
\cite{th96,in96}, but these works usually assume for simplicity 
that the dominant CR component consists of protons, and only some crude attempts 
have been done trying to generalise these predictions to heavier compositions
\cite{ma03}. 

Another contribution to the diffuse neutrino fluxes produced by CRs are 
those resulting from the interactions of the CRs present all over the 
Galaxy with the ambient gas in the inter stellar medium (ISM) \cite{st79,do93,
be93,in96}. The very low densities in the ISM ($n_{ISM}\simeq 1/{\rm cm}^3$) imply that 
essentially all mesons produced decay in this case without suffering any 
attenuation, and hence the neutrino fluxes are just the conventional ones 
from $\pi$, $K$ and $\mu$  decays, and moreover their spectral slopes follow the 
behaviour of the primary CR spectrum. These neutrino fluxes are proportional 
to the column density of the ISM along the particular direction considered, 
and  hence the flux is maximal in the direction to the Galactic Center, 
it is generally enhanced near the galactic plane while it is minimum in the 
direction orthogonal to it.

Along directions near the plane of the Milky Way, these diffuse fluxes of 
galactic neutrinos may overcome the atmospheric neutrino background 
at energies larger than $\sim 10^{14}$~eV (the actual energy 
depends on the model adopted for charm production and on the assumed
ISM column density, see below), 
but in the direction orthogonal to the plane they remain below the expected
flux from prompt atmospheric neutrinos from charmed particle decays up to the 
highest energies \cite{in96}. Here again the usual simplifying working
hypothesis is that the CR flux consists mainly of protons, and only
crude attempts have been done trying to generalise the predictions to
heavier compositions \cite{do93}.
Although some works have found a strong sensitivity 
to the assumed composition both for atmospheric and galactic CR
induced neutrinos, this issue is generally not treated appropriately 
and hence a more thorough analysis is required in order to estimate the neutrino 
fluxes with some confidence for $E_\nu>10^{14}$~eV.

The very high energy neutrino fluxes produced by CRs both in the atmosphere 
and in the Galaxy are one of the important targets of present and future 
neutrino telescopes, such as AMANDA, BAIKAL, ANTARES, ICECUBE, etc.. 
They are interesting per se, but they also represent the main background 
in the search of the fluxes arising from other potential neutrino sources, 
such as Active Galactic Nuclei (AGNs) \cite{st96} 
and Gamma Ray Bursts (GRBs) \cite{wa99}, which might give rise to detectable 
fluxes just in the energy range $10^{14}$--$10^{16}$~eV \cite{ah03}. 
It is hence of primary importance to establish with some confidence the 
actual value of the CR induced neutrino fluxes, and the aim of the present 
work is to point out that the predictions are quite sensitive to the detailed 
composition and to the behaviour of the spectrum of the individual nuclear 
components of the CRs with energies above those corresponding to the knee 
in the spectrum.

A major problem that one has to face here is that although the existence of 
the knee has been known for more than forty years, there is still no 
consensus  on the underlying physics responsible for this feature, and 
different proposed explanations for it can lead to quite different 
predictions for the behaviour of the CR composition and the individual 
spectra. On the observational side, due to the indirect nature of the 
measurements, which are based on the analysis of extensive air showers, 
also the situation is far from settled. 

A large class of possible scenarios to explain the knee are based on a 
rigidity dependent effect, such as a change on the CR acceleration 
efficiency at the sources \cite{fi86,jo86,ko00} or on a change in the 
escape mechanism of CRs from the Galaxy \cite{sy71,wd84,pt93,ca02a}, 
both of which, being magnetic effects, depend on the 
ratio $E_{CR}/Z$, where $Z$ is the CR charge.
Hence, in these scenarios the different nuclear components essentially 
steepen their spectra at an energy $E_Z\simeq ZE_{knee}$, and in this way 
the light components become more suppressed at smaller energies than the 
heavy ones. This leads to the prediction that the CR composition  should become 
increasingly heavier above $E_{knee}$, in agreement with the latest observations 
of KASCADE \cite{ka99} and the EAS-TOP/MACRO experiments \cite{ag03}.
In addition, the change in the spectral slope of each individual component 
usually turns out to be steeper than the change in slope of the total observed 
spectrum, which is $\Delta\alpha\simeq 3-2.7=0.3$. In particular, in the 
so-called diffusion/drift model \cite{pt93,ca02a,ca02b,ca03} 
in which the knee is due to a change in the escape mechanism of CRs 
from the Galaxy from one dominated by normal 
diffusion (with diffusion coefficient $D\propto E^{1/3}$) to one dominated 
by drift effects (with $D\propto E$), the spectral slope of each individual 
nuclear component changes by an amount $\Delta\alpha_Z\simeq 2/3$. 

As we will show below, these rigidity dependent scenarios have a profound 
impact on the predictions of high energy diffuse neutrino fluxes, and hence 
a clear understanding of the physics responsible for the knee and also 
better measurements of air showers at these energies are crucial to predict 
the expected neutrino fluxes and to extract conclusions from them once they 
will be measured.

\section{The cosmic ray spectrum}

Figure 1 shows the observed cosmic ray differential spectrum from several 
experiments \cite{na00}. As mentioned above, the total CR spectrum is well described
by power laws d$N_{CR}/{\rm d}E_{CR}\propto E_{CR}^{-\alpha}$, with the spectral
index changing from $\alpha\simeq 2.7$ to $\alpha\simeq 3$ at the
so-called knee, occurring
at $E_{knee}\simeq 3\times 10^{15}$~eV. The CR spectrum turns harder again at the 
ankle, corresponding to $E_{ankle}\simeq 5\times 10^{18}$~eV, a feature usually 
explained as the crossover between the dominance of the galactic CR component 
below the ankle and of the extragalactic one above it.

A simple way of describing the galactic CR flux within a rigidity dependent scenario
is to assume that each galactic component of charge $Z$ is given by
\begin{equation}
{{\rm d}N_Z\over{\rm d}E}\equiv \phi_Z=
{{\phi^<_Z\cdot \phi^>_Z}\over{\phi^<_Z+\phi^>_Z}}\ ,
\label{crfit1}
\end{equation}
where $\phi^<_Z$ ($\phi^>_Z$) is the CR flux below (above) the knee,
and is given by
\begin{equation}
\left\{ \begin{array}{cl}\phi^<_Z \\
\phi^>_Z\end{array}\right\}
=f_Z\phi_0\left(\frac{E}{E_0}\right)^{-\alpha_Z}\times
\left\{ \begin{array}{cl}1 \\
\left(E/ZE_k\right)^{-\Delta\alpha}\end{array}\right\}\ .
\label{crfit2}
\end{equation}
\begin{table}
\begin{tabular}{|r|r|r||r|r|r||r|r|r||r|r|r|} \hline
\multicolumn{1}{|c|}{Z}&\multicolumn{1}{|c|}{$f_Z$}&\multicolumn{1}{|c||}{$\alpha_Z$}&
\multicolumn{1}{|c|}{Z}&\multicolumn{1}{|c|}{$f_Z$}&\multicolumn{1}{|c||}{$\alpha_Z$}&
\multicolumn{1}{|c|}{Z}&\multicolumn{1}{|c|}{$f_Z$}&\multicolumn{1}{|c||}{$\alpha_Z$}&
\multicolumn{1}{|c|}{Z}&\multicolumn{1}{|c|}{$f_Z$}&\multicolumn{1}{|c|}{$\alpha_Z$}\\ \hline
 1 & 0.3775 & 2.71 & 8 & 0.0679 & 2.68 & 15 & 0.0012 & 2.69 & 22 & 0.0049 & 2.61 \\
 2 & 0.2469 & 2.64 & 9 & 0.0014 & 2.69 & 16 & 0.0099 & 2.55 & 23 & 0.0027 & 2.63 \\
 3 & 0.0090 & 2.54 & 10& 0.0199 & 2.64 & 17 & 0.0013 & 2.68 & 24 & 0.0059 & 2.67 \\
 4 & 0.0020 & 2.75 & 11& 0.0033 & 2.66 & 18 & 0.0036 & 2.64 & 25 & 0.0058 & 2.46 \\
 5 & 0.0039 & 2.95 & 12& 0.0346 & 2.64 & 19 & 0.0023 & 2.65 & 26 & 0.0882 & 2.59 \\
 6 & 0.0458 & 2.66 & 13& 0.0050 & 2.66 & 20 & 0.0064 & 2.70 & 27 & 0.0003 & 2.72 \\
 7 & 0.0102 & 2.72 & 14& 0.0344 & 2.75 & 21 & 0.0013 & 2.64 & 28 & 0.0043 & 2.51 \\ \hline
\end{tabular}
\caption{Cosmic ray fractional abundances (for $E=1$~TeV) and low energy 
spectral indices from hydrogen to nickel.}
\label{Table 1}
\end{table}
In this expression,  $\phi_0=3.5\times 10^{-13} {\rm m^{-2}s^{-1}sr^{-1}eV^{-1}}$
is the total CR flux at the reference energy $E_0$, hereafter adopted as
$E_0=1$~TeV, $f_Z$ is the fractional CR abundance 
 at the same energy, $\alpha_Z$ is the low energy spectral
index, $\Delta\alpha$ is the spectral index change across the knee (which is the
same for all CR components) and $E_k$ is a parameter that fixes the position 
of the knee, which is conveniently chosen to fit the observations.
The expression in Eq.~(\ref{crfit1}) provides then a smooth
interpolation\footnote{One may eventually generalise the expression in
  Eq.~(\ref{crfit1}) as
$\phi=\phi^<\cdot \phi^>/(\phi^{<n}+\phi^{>n})^{1/n}$, where the parameter
  $n$ will control the `width' of the knee, with $n\gg 1$ leading
  to an abrupt change in slope. We find however that $n=1$ provides a
  very satisfactory fit to the CR data for the $\Delta\alpha=2/3$ case considered.} 
between the flux $\phi^<_Z$ at low energies $(E<ZE_k)$ and $\phi^>_Z$ at
high energies $(E>ZE_k)$.
\begin{figure}[t]
\centerline{{\epsfxsize=5.truein\epsfysize=3.5truein\epsffile{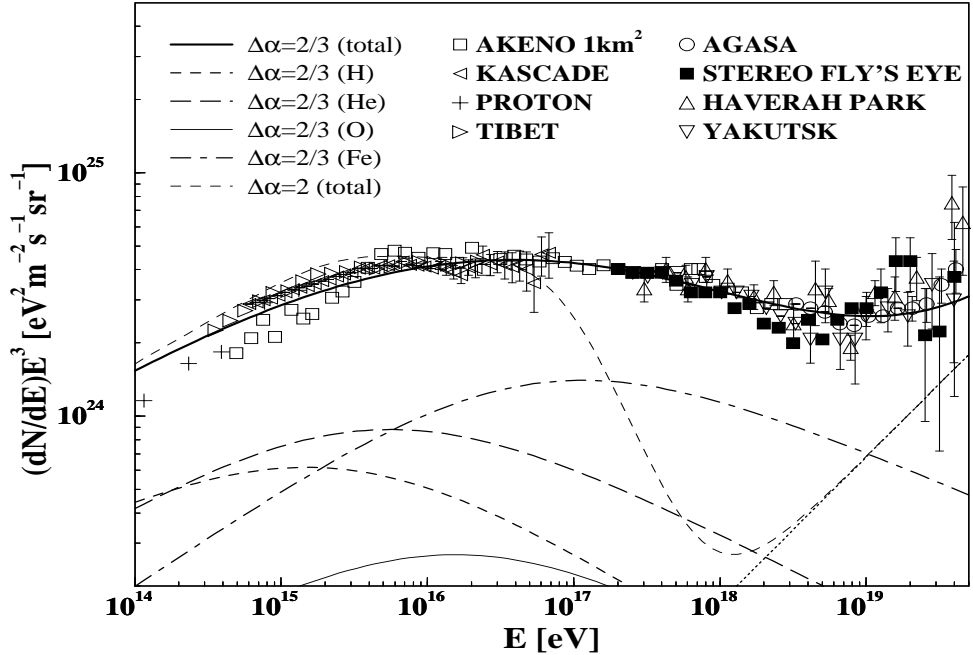}}}
\caption{CR spectra obtained from considering rigidity-dependent scenarios 
parametrized according to equations (\ref{crfit1}) and (\ref{crfit2}). For the spectral
index change $\Delta\alpha=2/3$, the main contributions to the total flux, 
which correspond to nuclei of H, He, O and Fe, are also indicated. The dotted
straight line corresponds to the extragalactic flux given by equation (\ref{xgflux}).   
Also shown are the relevant experimental observations.}
\label{fig1}
\end{figure}       
The CR abundances at $E=1$~TeV and the low-energy spectral indices 
are shown in Table 1, and were taken from the
data compiled in \cite{wi98,ho03}. 

For the extragalactic component, we will assume for definiteness 
that it consists of protons (this would anyhow only affect the
predictions for $E_\nu>10^{17}$~eV) and that it is given by 
\begin{equation}
\left({{\rm d}N\over{\rm d}E}\right)_{XG}=
6.8\times 10^{-34}\left({{E}\over{10^{19}\ {\rm eV}}}\right)^{-2.4}
{\rm m^{-2}s^{-1}sr^{-1}eV^{-1}}\ ,   
\label{xgflux}
\end{equation}
i.e. a flux similar to that considered in \cite{ca03}.   

For the purpose of illustration, we display in Figure 1 the cases of both 
$\Delta\alpha=2/3$, corresponding to the diffusion/drift model 
(when a Kolmogorov spectrum 
of fluctuations is assumed for the random magnetic field component 
in which CR particles propagate
\cite{ca02a,ca02b,ca03}), and $\Delta\alpha=2$, corresponding to an
extreme model advocated in \cite{ho03}.
Figure 1 shows the fits that result from taking $\Delta\alpha=2/3$ 
with the parameter value
$E_k=2.2\times 10^{15}$eV, and $\Delta\alpha=2$ with $E_k=3.1\times 10^{15}$eV.
While the former exhibits an excellent agreement with the experimental
observations, it can be seen that the latter shows an abrupt suppression above
$\sim 10^{17}$~eV which results from the steep suppression of the different CR
components above their knees \cite{ho03}, and due to this we will
hereafter just focus
on the $\Delta\alpha=2/3$ case. 
   
\section{The flux of atmospheric neutrinos}
The CR particles reaching the top of the atmosphere produce secondary fluxes
of hadrons and leptons that are usually described by means of coupled cascade
equations \cite{ga90,li93,th96,pa99}, which can be solved analytically under 
appropriate symplifying assumptions. 
In this work we will compute the fluxes of muon neutrinos and 
antineutrinos, which are the ones more readily detectable at
underground detectors. At the energies considered the neutrino flavor 
oscillations will have no effect on the atmospheric neutrinos,
although they may be relevant for the neutrinos produced by CR
interactions with the ISM (eventually redistributing in that case the
produced $\nu_e$ and $\nu_\mu$ fluxes evenly among the three flavors, 
something that can be incorporated into the final results straightforwardly).

The main contributions to the neutrino flux in the range of interest of
this work (namely, for $E=10^3-10^8$~GeV) arise, on the one hand, 
from the decay of the mesons $\pi^\pm, K^\pm, K_S^0$ and $K_L^0$, 
which produce the so called conventional atmospheric neutrinos that dominate
at low energies, and on the other hand, from the decay of the charmed 
mesons $D^0,D^\pm,\Lambda_c$ and $D_s$, which give rise to the so called 
prompt charm neutrino flux that dominates at very high energies. 
The muon contribution to the atmospheric neutrino flux is 
completely negligible in this context, since above $E\sim 10^3$ GeV
the muons reach the ground and are stopped before decaying.  

\subsection{Nucleon and meson fluxes}

The transport equation associated to a given cascade component $j$ can
be written as
\begin{equation}
{{{\rm{d}}\phi_j}\over{{\rm{d}}X}}(E,X)=-{{\phi_j}\over{\lambda_j}}(E,X)-
{{\phi_j}\over{\lambda_j^d}}(E,X)+\sum_kS_{kj}(E,X)\ ,
\label{casc}
\end{equation}      
where as usual $X$ is the slant depth, $\lambda_j$ is the interaction length
in air and $\lambda_j^d$ is the decay length. Both characteristic lengths are 
measured in g/cm$^2$ units (i.e. they include the factor $\rho(X)$ that 
corresponds to the local density of the atmosphere). Also notice that the decay
length is linear in the energy due to the Lorentz time dilation factor, while
the interaction length exhibits instead a much milder energy dependence.   
The last term in equation (\ref{casc}) corresponds to the production/regeneration term
given by
\begin{equation}
S_{kj}(E,X)=\int_E^\infty{\rm{d}}E'{{\phi_k(E',X)}\over{\lambda_k(E')}}
{{1}\over{\sigma_{kA}(E)}}{{{\rm{d}}\sigma_{kA\to j}(E,E')}\over{{\rm{d}}E}}\ ,
\label{coupling1}
\end{equation}
which actually couples the transport equations of different 
cascade components. 
Notice that $\sigma_{kA}$ here refers to the total $k+$air cross section,
while $\sigma_{kA\to j}$ corresponds to the process $k+$air $\to$ $j+$
anything.
Since the energies involved in the nuclear 
collisions between CRs and atmospheric nuclei are much higher than nuclear binding
energies, only nucleon-nucleon interactions are relevant in this context. 
 The different cascade 
components that we will take into account are hence $N$ (that 
corresponds to the nucleon 
component in which protons and neutrons are grouped together)
and $M=\pi^\pm, K^\pm, K_L^0,D^0,D^\pm,\Lambda_c$ and $D_s$. 
As in \cite{li93}, the $K_S^0$ can be considered as contributing to the pion flux. 
Moreover, we will assume for simplicity that the only non-negligible
terms will be those corresponding to regeneration (i.e. $S_{NN}$ and $S_{MM}$) and to 
meson production by nucleons (i.e. $S_{NM}$) \cite{li93,th96}. 

Let us consider in particular fluxes with the form $\phi_k(E,X)= E^{-\beta_k}g(X)$ 
(i.e. we consider the separation of variables in the fluxes, 
in which the energy dependent factor has a power-law behavior). Then, it turns out that 
\begin{equation}
S_{kj}(E,X)={{\phi_k(E,X)}\over{\lambda_k(E)}}Z_{kj}(E)\ ,
\label{coupling2}
\end{equation} 
where $Z_{kj}$ is the production/regeneration moment defined by
\begin{equation}
Z_{kj}(E)=\int_0^1{\rm{d}}x\ x^{\beta_k-1}{{1}\over{\sigma_{kA}(E)}} 
{{{\rm{d}}\sigma_{kA\to j}(E,x)}\over{{\rm{d}}x}}\ ,
\label{zmoment}
\end{equation}
with $x\equiv E/E'$.

Assuming that the interaction lengths are energy independent
and that the differential production distribution is Feynman scaling, the
production/regeneration moments themselves become energy independent. Indeed, for the 
relevant interaction lengths and production/regeneration moments involving 
nucleons and non-charmed mesons, we consider the energy independent values 
given in \cite{ga90,li93}.
When necessary, we also take the $\beta_k-$dependence from fits to the data compiled
and plotted in \cite{ga90}. 
Moreover, following \cite{pa99}, the charm interaction lengths
and regeneration moments are all set equal to the kaon ones.   

For the production of charmed particles, however, one has to take into account
the energy dependent moment given by equation (\ref{zmoment}), since the results are
very sensitive to the particular cross section adopted for the 
charm production process. Indeed, the various estimates of the prompt 
atmospheric fluxes calculated so far are found to differ by almost 2 orders of
magnitude \cite{ge03}.
   
Due to their relatively large mass, charm quarks are usually considered to be 
produced in hard processes which can be well described by perturbative QCD (pQCD).
To leading order (LO) in the coupling constant, the processes that contribute to 
the charm production cross section are the gluon-gluon fusion process 
$gg\to c\overline{c}$ and the quark-antiquark annihilation process 
$q\overline{q}\to c\overline{c}$. However, at the next to leading order (NLO) the
gluon scattering process $gg\to gg$ shows up, giving a very significant contribution to 
the total cross section which increases it by  a factor of $\sim
2-2.5$ \cite{na88,be89,ma89}. Concerning higher
order corrections, their contribution is certainly
small since there are no qualitatively new channels opening up.
Several parton distribution functions have been proposed and provide theoretical
pQCD predictions that fit the available accelerator data reasonably well. However,
in order to calculate the atmospheric charm flux the parton distribution functions
need to be extrapolated to very small parton momentum fractions, typically
corresponding to $x\simeq 4$~GeV/$E$ \cite{ma03}, 
which at the highest energies is well outside of the measured 
region ($x>10^{-5})$, and the uncertainty in the extrapolation 
affects significantly the final results \cite{pa99,ge03}.            
In order to illustrate this uncertainty range, we display the results
obtained with two different structure distribution functions, 
namely the CTEQ3 parton distribution function \cite{la95} and the Golec-Biernat, 
W\"usthoff (GBW) model \cite{go99}, which includes gluon saturation effects.  
The corresponding charm production cross sections were obtained either directly
from the fit given in \cite{ma03} (for the results based on the GBW saturation 
model), or by fitting the results given in figure~4 of \cite{pa99} 
for the CTEQ3 structure functions
set with $M=2m_c=2\mu$, with the charm mass $m_c=1.3$~GeV, and
interpolating the results for different energies.    
For the nucleon-air cross section, we used the parametrization given by \cite{mi94}
\begin{equation}
\sigma_{NA}(E)=\left[280-8.7\ln\left({\frac{E}{\rm{GeV}}}\right)+
1.14\ln^2\left({\frac{E}{\rm{GeV}}}\right)\right]{\rm{mb}}\ .
\label{sigmaN}
\end{equation}

Once all relevant interaction lengths and production/regeneration 
moments are determined, 
the nucleon and meson atmospheric fluxes can be calculated from the coupled cascade
equations given by (\ref{casc}) and (\ref{coupling2}). 
Recalling that these expressions were obtained under the assumption of 
power-law nucleon and meson spectra, we will first discuss the results 
for an initial nucleon flux given according to 
$\phi_N(E,X=0)=\phi_{0N}E^{-\gamma}$, discussing the general case
further below.

Under the assumptions mentioned above, the nucleon flux develops independently 
from the meson fluxes and is given by
\begin{equation}
\phi_N(E,X)=e^{-X/\Lambda_N}\phi_{0N}E^{-\gamma}\ ,
\label{nucleons}
\end{equation}  
where the nucleon attenuation length is defined as 
\begin{equation}
\Lambda_N={{\lambda_N}\over{1-Z_{NN}}}\ .
\label{LambdaN}
\end{equation} 
 
Concerning the meson cascade equations, they are usually solved by considering 
separately the low energy solution $\phi_M^L$ 
(which neglects the interaction and regeneration terms, 
since $\lambda_M\gg\lambda_M^d$ at low energies) 
and the high energy solution $\phi_M^H$ (which neglects instead the decay term, 
since $\lambda_M\ll\lambda_M^d$ at very high energies) \cite{ga90,li93,th96}.
 
The high energy solution is given by
\begin{equation}
\phi_M^H(E,X)={{Z_{NM}}\over{1-Z_{NN}}}{{e^{-X/\Lambda_M}-e^{-X/\Lambda_N}}
\over{1-\Lambda_N/\Lambda_M}}\phi_{0N}E^{-\gamma}\ ,
\label{mesonsH}
\end{equation} 
where the meson attenuation length $\Lambda_M$ is defined analogously to equation 
(\ref{LambdaN}). 

For the low energy case, the solution is obtained by replacing $\Lambda_M\to\lambda_M^d$ 
in the last equation. Recalling that $\lambda_M^d\propto E$, the resulting expression
couples the dependence on the energy and the slant depth, and hence it
does not lead to just 
a simple power law energy spectrum.
However, for not very small values of $X$ the solution reduces to 
\begin{equation}
\phi_M^L(E,X)={{Z_{NM}}\over{1-Z_{NN}}}{{\lambda_M^d}\over{\Lambda_N}}
e^{-X/\Lambda_N}\phi_{0N}E^{-\gamma}\ .
\label{mesonsL}
\end{equation}

In equations (\ref{nucleons})--(\ref{mesonsL}), 
$Z_{NN},Z_{NM},$ and $Z_{MM}$ are to be evaluated for the nucleon spectral index 
$\gamma$. Notice also that the high energy meson flux has the same spectral index as the nucleon flux, 
while the low energy meson solution is flatter by an extra power of
the energy, due to the energy dependence implicit in $\lambda_M^d$. 

\subsection{Neutrino flux}
The neutrino flux produced via the meson weak decay can be calculated following
the same scheme described above for the atmospheric nucleon and meson fluxes. Indeed,
the neutrino flux is given by a transport equation analogous to equation (\ref{casc}), 
but actually much simpler since it contains only the source terms arising from meson decay, i.e.
\begin{equation}
{{{\rm{d}}\phi_\nu}\over{{\rm{d}}X}}(E,X)=\sum_MS_{M\nu}(E,X)\ ,
\label{nuflux1}
\end{equation}      
where 
\begin{equation}
S_{M\nu}(E,X)=\int_E^\infty{\rm{d}}E'{{\phi_M(E',X)}\over{\lambda_M^d(E')}}
{{1}\over{\Gamma_M(E)}}{{{\rm{d}}\Gamma_{M\nu}(E,E')}\over{{\rm{d}}E}}\ .
\label{nuflux2}
\end{equation}
The decay distribution can be put in terms of $F_{M\nu}$, the inclusive neutrino spectrum 
in the decay of meson $M$, by means of the relation 
\begin{equation}  
{{1}\over{\Gamma_M(E)}}{{{\rm{d}}\Gamma_{M\nu}(E,E')}\over{{\rm{d}}E}}=B_{M\nu}F_{M\nu}(E,E')\ ,
\end{equation}  
where $B_{M\nu}$ is the branching ratio for the decay of meson $M$ 
into a state with the given neutrino $\nu$.
In the ultrarelativistic limit, the inclusive neutrino spectrum scales as 
$F_{M\nu}(E,E')=F_{M\nu}(E/E')/E'$ \cite{ga90,li93}. Moreover, we have already derived
the asymptotic solutions for the meson fluxes, which are of the form 
$\phi_M(E,X)\propto E^{-\beta}g(X)$, with $\beta=\gamma$ ($\beta=\gamma-1$) for the high (low)
energy fluxes (see equations (\ref{mesonsH}) and (\ref{mesonsL})).
Thus, the source terms given by equation (\ref{nuflux2}) can be rewritten as
\begin{equation}
S_{M\nu}(E,X)={{\phi_M(E,X)}\over{\lambda^d_M(E)}}Z_{M\nu}^{\beta+1}(E)\ ,
\label{nuflux3}
\end{equation}  
where the $\beta-$dependent meson decay moments are defined 
by\footnote{The branching ratio is here included in the definition of the decay
moments, in analogy with the production/regeneration moments that include
the multiplicity of the final states. This coincides with \cite{th96} but
differs from \cite{li93}.}
\begin{equation}
Z_{M\nu}^\beta(E)=B_{M\nu}\int_0^1{\rm{d}}x\ x^{\beta-1} F_{M\nu}(x)\ .
\label{zdec}
\end{equation}
The decay moments used in this work were calculated by fitting the $\beta-$
dependence of the relevant moments tabulated in \cite{th96} for $2.7\leq\beta\leq 4$,
which were determined with Lund Monte Carlo simulation programs. 

Then, the asymptotic low and high energy neutrino fluxes can now be determined 
from the equations (\ref{nuflux1}) and (\ref{nuflux3}), which involve the
meson decay moments and the corresponding high and low energy meson fluxes
derived in equations (\ref{mesonsH}) and (\ref{mesonsL}). The low energy solution      
reads
\begin{equation}
\phi_\nu^L(E,X)=Z_{M\nu}^\gamma{{Z_{NM}}\over{1-Z_{NN}}}
\left(1-e^{-X/\Lambda_N}\right)\phi_{0N}E^{-\gamma}\ .
\label{atmosL1}
\end{equation}
As expected, the neutrino flux develops rapidly, on the scale of a
nucleon interaction length,  and then remains stable. Hence, 
at ground level ($X\gg\Lambda_N$) the low energy flux will be
\begin{equation}
\phi_\nu^L(E)=Z_{M\nu}^\gamma{{Z_{NM}}\over{1-Z_{NN}}}\phi_{0N}E^{-\gamma}\ .
\label{atmosL2}
\end{equation}
Similarly, the high energy flux at ground is
\begin{equation}
\phi_\nu^H(E)=Z_{M\nu}^{\gamma+1}{{Z_{NM}}\over{1-Z_{NN}}}{{\ln\left(\Lambda_M/
\Lambda_N\right)}\over{1-\Lambda_N/\Lambda_M}}{{\epsilon_M}\over{\cos\theta}}
\phi_{0N}E^{-(\gamma+1)}\ ,
\label{atmosH}
\end{equation}
where $\theta$ is the zenith angle of the CR incidence direction
considered and 
\begin{equation}
\epsilon_M={{m_Mch_0}\over{\tau_M}}\ ,
\end{equation}
with $m_M$ and $\tau_M$ the meson's mass and mean life, 
respectively, and $h_0=6.4$~km a typical
scale height for density variations in the atmosphere. 
Since the production/regeneration moments
$Z_{NN},Z_{NM},$ and $Z_{MM}$ are the same as those appearing in the 
meson fluxes, they 
are to be evaluated for the nucleon spectral index $\gamma$. 
Equation (\ref{atmosH}) is actually valid for zenith angles
sufficiently small so that the curvature of 
the earth can be neglected, and assumes a density profile of an
isothermal  atmosphere 
(i.e. $\rho(h)\propto\exp(-h/h_0)$), 
which is appropriate in the stratosphere ($h\geq 11$~km) 
where most particle interactions occur \cite{li93,th96}. 
For very large angles, 
one can still use equation (\ref{atmosH}) with the simple
prescription of replacing $\theta\to\theta^*(\theta,h=30$~km), 
where  for a given line of sight that corresponds to an angle
$\theta$ at the observer's position, the angle $\theta^*(\theta,h)$ is
the zenith angle that would be observed at a location where this same
line of sight is at a height $h$ with respect to the ground \cite{li93}. 
For instance, for CRs incident in the horizontal direction
($\theta=90^\circ$), one has that 
$\theta^*\simeq \arcsin(1+h/R_\oplus)^{-1}\simeq 84.5^\circ$.     

In order to obtain a general expression for the atmospheric neutrino fluxes
deep in the atmosphere, the asymptotic solutions given by equations 
(\ref{atmosL2}) and (\ref{atmosH}) can be joined together by means of the 
interpolation function 
\begin{equation}
\phi_\nu={{\phi^L_\nu\cdot \phi^H_\nu}\over{\phi^L_\nu+\phi^H_\nu}}\ ,
\label{interpol}
\end{equation}
which is actually analogous to the interpolation function given in equation (\ref{crfit1})
to join the fluxes associated to each galactic CR component below and
above their knees.
 
We described so far a calculational scheme that provides the nucleon, meson and
neutrino fluxes produced by an initial power law nucleon flux $\phi_N(E,X=0)=\phi_{0N}E^{-\gamma}$ 
hitting the top of the atmosphere. We now seek the corresponding results for the full 
CR spectrum arriving at the earth, taking into account both the CR
composition and the break in the CR spectra at the knee. 
Let $\phi_Z$ be the CR flux associated to the  
CR component of nuclei of charge $Z$ and mass $Am_N$ (with $A$ being the average mass
number corresponding to element $Z$). This nuclear component provides a nucleon 
flux $\phi_{N,Z}$ given by $\phi_{N,Z}(E_N)=A\phi_Z(E=AE_N)$. If the flux of the CR component 
of charge $Z$ is $\phi_Z(E)=\phi_{0Z}(E/E_0)^{-\gamma_Z}$, 
the corresponding nucleon flux is then given by  
\begin{equation}
\phi_N(E)=\sum_Z A^{2-\gamma_Z}\phi_{0Z}\left({E\over E_0}\right)^{-\gamma_Z}\ .
\label{crnucleon}
\end{equation}
>From these considerations, the procedure to calculate the neutrino flux produced by the 
galactic CR flux parametrized by equations (\ref{crfit1}) and (\ref{crfit2}) is now
straightforward. Indeed, the CR component of charge $Z$ is represented by two power
law fluxes $\phi_{Z}^<$ and $\phi_{Z}^>$, with spectral indices $\alpha_Z$ and    
$\alpha_Z+\Delta\alpha$, respectively. For each power law flux, we can compute the 
corresponding nucleon flux given by equation (\ref{crnucleon}) and use it to 
calculate the associated neutrino flux. Then, the final neutrino flux produced by
the given CR component can be obtained by interpolating the two solutions, using 
again the interpolation function analogous to equation~(\ref{crfit1}).
Finally, the total neutrino flux results from summing over the contribution of all
galactic CR components together with the additional contribution 
from the extragalactic component given by equation~(\ref{xgflux}).

Figure 2 shows the atmospheric neutrino fluxes that correspond to 
the $\Delta\alpha=2/3$
rigidity dependent scenario, with the parametrization formulae and 
parameter values given
in section 2. The figure shows the contribution coming from different 
components and for
different cases, namely the vertical and horizontal conventional, 
prompt charm/GBW and
prompt charm/CTEQ3 fluxes. For the prompt charm flux calculated with 
the GBW model, the
figure also shows the total vertical and horizontal neutrino fluxes. 
Finally, in the latter
case the conventional, charm and total neutrino fluxes produced by 
the extragalactic CR component is also indicated. 
\begin{figure}[t]
\centerline{{\epsfxsize=4.7truein\epsfysize=3.2truein\epsffile{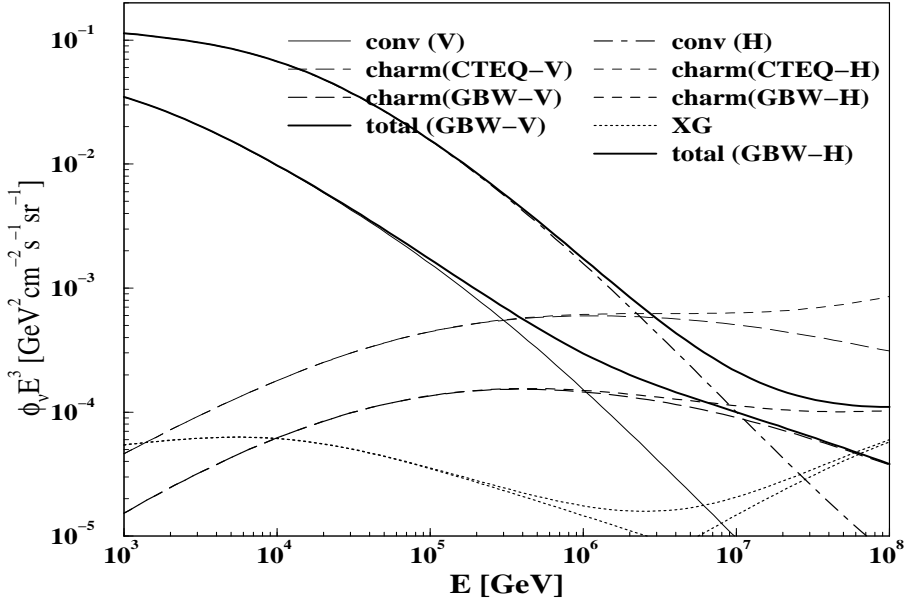}}}
\caption{Atmospheric ($\nu_\mu+\bar\nu_\mu$) fluxes for the $\Delta\alpha=2/3$
rigidity dependent scenario (that corresponds to the CR spectrum plotted in
figure 1). The vertical and horizontal conventional, prompt charm/GBW and
prompt charm/CTEQ3 fluxes are shown, as well as the total vertical and horizontal 
neutrino fluxes for prompt charm/GBW. In the latter case, the conventional, charm and 
total neutrino fluxes produced by the extragalactic CR 
component is also indicated.}
\label{fig2}
\end{figure} 
As is clear from equation (\ref{crnucleon}), the induced neutrino fluxes are
sensitive, on the one hand, to the CR composition assumed and, on the other hand, 
to the model assumed to reproduce the steepening of the CR spectrum at the knee. 
Previous works that have taken the knee into account have assumed for simplicity 
that the dominant CR component consists of protons alone, having a
spectral change $\Delta\alpha=0.3$ at $E_{knee}$. In order to observe the 
effect of the CR composition on the neutrino fluxes, figure 3 compares the total
horizontal atmospheric fluxes (with the prompt charm component calculated with the GBW
model) as obtained for different assumptions on the CR spectrum, namely the $\Delta\alpha=2/3$
rigidity dependent scenario (that corresponds to the results given already in figure 2), 
the same total CR spectrum but assuming it consists only of protons, and the CR 
spectrum used in \cite{th96,in96} (which is assumed to be constituted by protons and has a single 
spectral index change of $\Delta\alpha=0.3$ at $E_{knee}=5\times 10^{15}$eV).     
The latter CR spectrum has a normalization which is lower by a factor of $\sim 2.6$ compared to the 
former ones, and indeed it seems to lay somewhat below the present observational data.       
Comparing the results for the same total CR spectrum, one can observe that the neutrino flux 
corresponding to the composition of different nuclear species is clearly below that produced
by a CR spectrum formed only by protons, and also that this effect is more significant at 
higher energies. Indeed, equation (\ref{crnucleon}) shows that 
the nuclear CR component of charge $Z$ and mass $A$ gives a contribution to the nucleon flux 
which is suppressed by a factor of $A^{2-\alpha_Z}\sim A^{-0.7}$ below the respective
knee, while it becomes suppressed by a factor 
$A^{2-\alpha_Z-\Delta\alpha}\sim A^{-1.4}$ well above it. This effect was not appropriately 
treated in \cite{ma03}, where the suppression was considered to be
given by a factor of $A^{-2}$, and moreover the impact of the change in the
composition at  the knee was never discussed before. The strong
suppression of the $\nu$ flux produced by the heavier components
implies that the light (H and He) components are still responsible for
a large fraction of the neutrinos above their knee, and hence the
change in slope of the individual components is also reflected in the
change in slope of the neutrino fluxes.

\begin{figure}[t]
\centerline{{\epsfxsize=4.5truein\epsfysize=3.1truein\epsffile{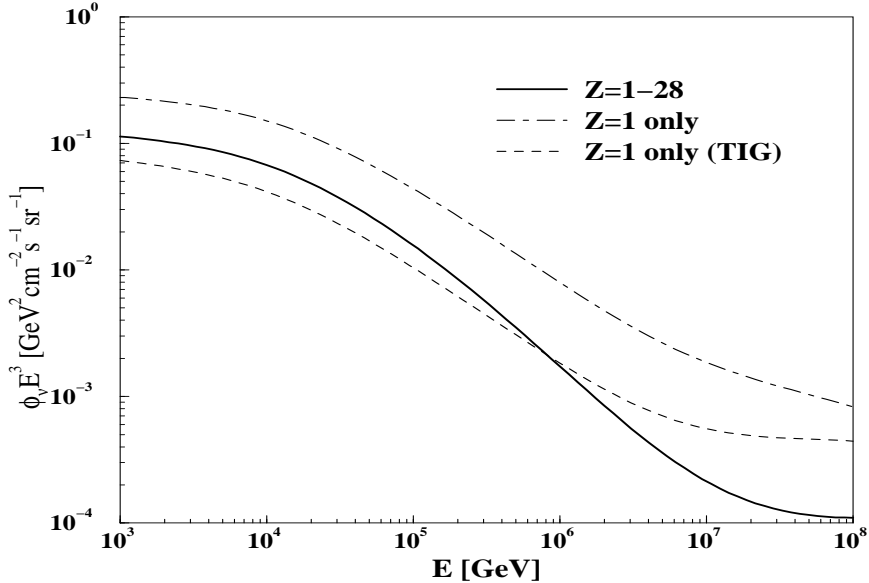}}}
\caption{Comparison of the total horizontal atmospheric ($\nu_\mu+\bar\nu_\mu$) 
flux for prompt charm/GBW produced
by CRs under different assumptions concerning spectrum and composition: 
the $\Delta\alpha=2/3$ rigidity dependent scenario that considers the contribution of CR 
components with charge $1\leq Z\leq 28$; the same total CR spectrum but assuming it consists 
only of protons; and the CR spectrum used in \cite{th96,in96} (TIG), which also takes only protons
into account but has a lower normalization.}
\label{fig3}
\end{figure}   

\section{The flux of galactic neutrinos}
The neutrino fluxes produced during the propagation of the CR particles through the
ISM can be determined following the procedure described in the preceding 
section for the production of atmospheric neutrinos. 
For simplicity, we consider that the CRs are distributed homogeneously 
within the galactic disk.  
The highly relativistic CR particles interact with the ambient gas
in the ISM, which is a very low-density, non-relativistic plasma constituted chiefly by 
atomic and molecular hydrogen that can be described by an homogeneous
nucleon density $n_{ISM}\simeq 1/{\rm cm}^3$ \cite{be90}. 
The treatment follows essentially the same considerations described above, but 
in this case one has to consider that the interaction of the secondaries produced in the
nucleon-nucleon interactions is completely negligible due to the
extremely low ISM densities, which allow for their decay well before
interactions can take place, irrespective of the energy considered. 
Hence, the main contributions to the diffuse neutrino flux produced
in the Galaxy arise from the decay of pions and muons, while the additional contributions 
coming from production and decay of heavier mesons can be safely disregarded. 

As in the preceding section, let us first consider an initial nucleon flux given by
$\phi_N(E,X=0)=\phi_{0N}E^{-\gamma}$. Notice that in this context the atmospheric slant depth
should be replaced by the ISM column density traversed along the line of sight (i.e. 
$X\to Rm_Nn_{ISM}$, where $R$ is the distance measured from the border of the galactic disk
along the line of sight). Due to the extreme faintness of the ISM, we should now take into account
that\footnote{In this case, $\lambda_N$ corresponds to the interaction
length for nucleons propagating through the ISM. Notice that
$\lambda_N$, expressed in g/cm$^2$, is actually very similar to the one
corresponding to nucleon propagation in air. In contrast, the meson
decay length $\lambda_M^d$, which 
is linear in the mass density of the medium, will be much smaller
in the ISM than in the atmosphere.} $X\ll\lambda_N$. Indeed, it turns out that
\begin{equation}
{{X}\over{\lambda_N}}=3.1\times 10^{-6}\left({{R}\over{\rm{kpc}}}\right)
\left({{\sigma_{NN}}\over{\rm{mb}}}\right)\ , 
\end{equation}  
where the nucleon-nucleon total cross section is 
\begin{equation}
\sigma_{NN}\simeq\left[35.49+0.307\ln^2\left(s/28.94\
  \rm{GeV}^2\right)\right]\ {\rm mb}\ ,
\label{sigmaNN}
\end{equation}
and where $s\simeq 2m_NE$ stands for the center of mass energy squared \cite{pdg}.
 
Hence, considering equation (\ref{atmosL1}) with $X\ll\lambda_N$, the neutrino flux 
produced from pion decay in the ISM is  
\begin{equation}
\phi_\nu(E,X)=Z_{M\nu}^\gamma Z_{NM}{{X}\over{\lambda_N}}\phi_{0N}E^{-\gamma}\ .
\label{gal}
\end{equation} 

In order to determine the muon decay contribution, it suffices to estimate it  
directly from the results obtained already for pion decay. 
From the decay kinematics, the mean fraction of energy (relative to the parent particle)
in the $\pi\to\mu+\nu_\mu$ decay is $K_\pi=0.21$ for the $\nu_\mu$ and 0.79 for the
muon \cite{vo80}. Analogously, in the $\mu\to e+\nu_e+\nu_\mu$ decay the effective 
fraction of energy for the resulting $\nu_\mu$ is 0.35 \cite{vo80}, i.e. a fraction 
$K_\mu=0.28$ relative to the original pion. 
Hence, one expects a muon decay contribution approximately a factor 
$(K_\mu/K_\pi)^\gamma$ larger than the neutrino flux from pion decay 
(for instance, a factor of $2.1$ for a nucleon spectral index $\gamma=2.7$, in sensible
agreement with more detailed calculations \cite{in96}).  

\begin{figure}[th!]
\centerline{{\epsfxsize=4.1truein\epsfysize=2.6truein\epsffile{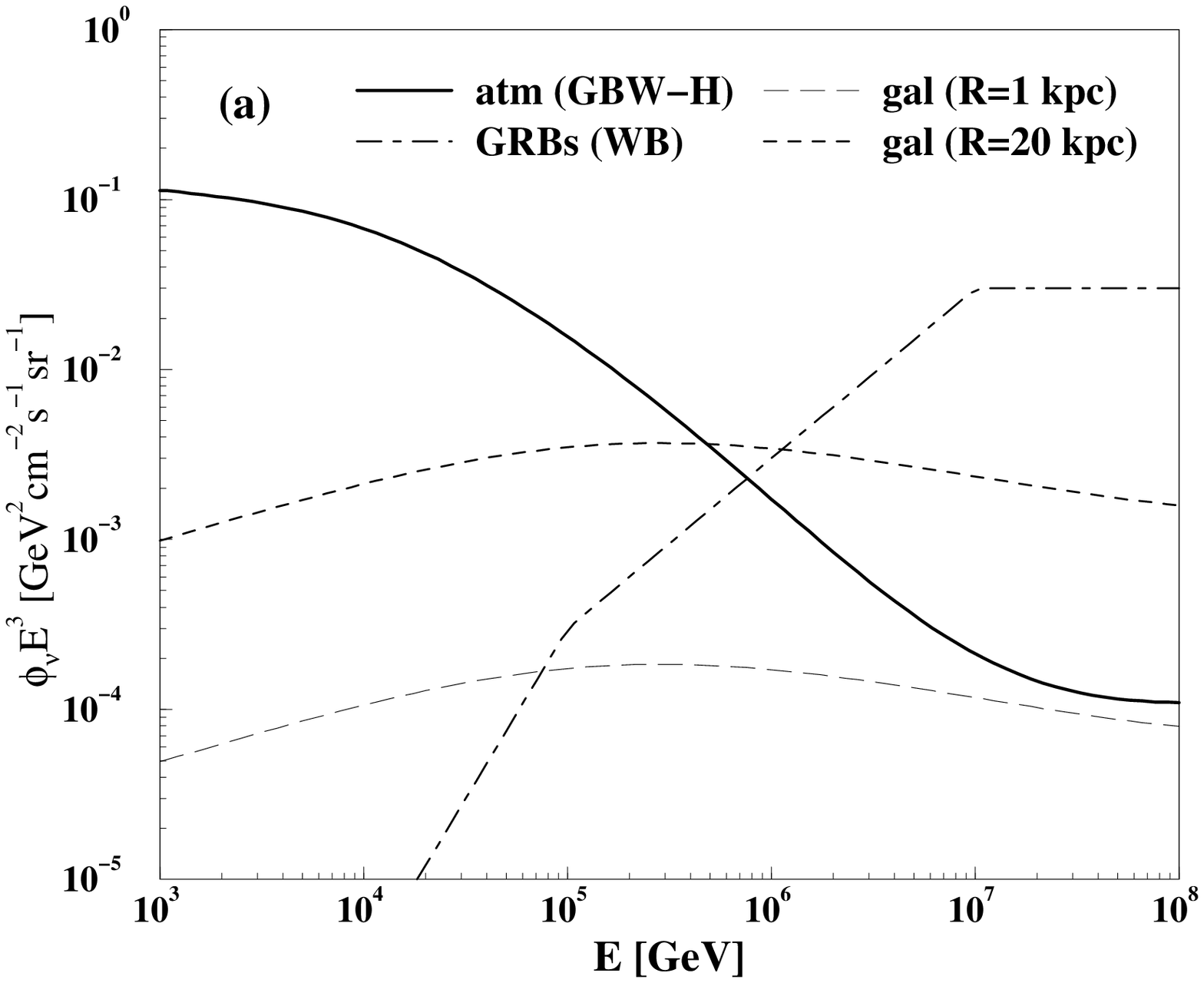}}}
\centerline{{\epsfxsize=4.1truein\epsfysize=2.6truein\epsffile{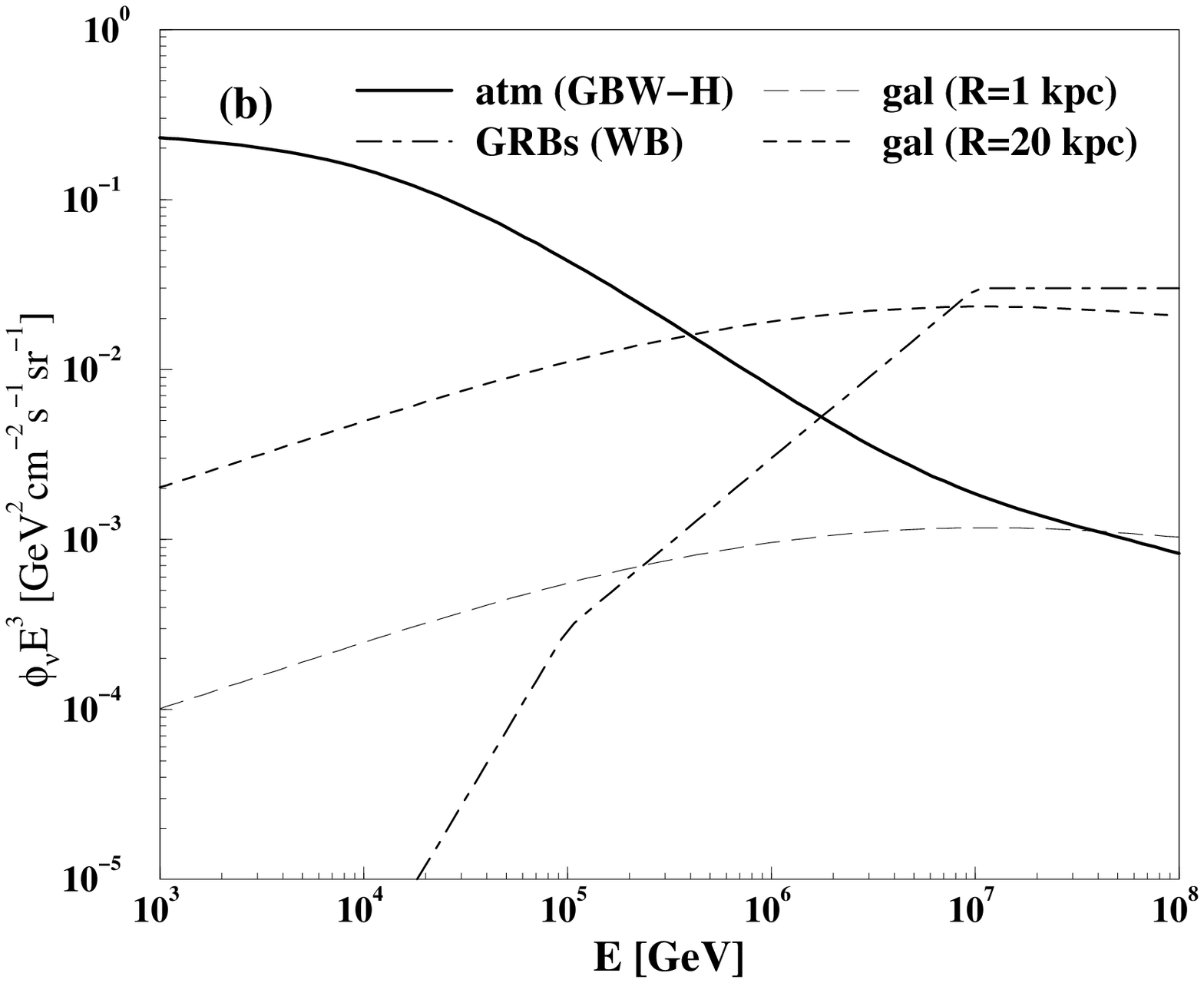}}}
\caption{Neutrino fluxes ($\nu_\mu+\bar\nu_\mu$) 
produced by CRs in the ISM (for different path lengths traversed through the ISM, 
namely $R=1$ and 20 kpc) and in the atmosphere (for the horizontal direction, using the GBW
model for the prompt charm contribution) assuming different CR compositions:
(a) the $\Delta\alpha=2/3$ rigidity dependent scenario that considers the contribution of CR 
components with $1\leq Z\leq 28$; (b) the same total CR spectrum but assuming it consists 
only of protons. For comparison, also shown are the Waxman-Bahcall predictions for the 
astrophysical diffuse neutrino fluxes produced in their model of GRBs.}
\label{fig4}
\end{figure}  
As commented above, the galactic neutrino flavor oscillations may redistribute the produced 
$\nu_\mu$ and $\nu_e$ fluxes evenly among the three flavors. In that case, 
the calculation of the electron neutrino flux produced by CR interactions with the ISM
is also required in order to determine the galactic muon
neutrino flux reaching the earth. This flux is mainly produced in the $\mu\to e+\nu_e+\nu_\mu$
decay, in which the effective fraction of energy for the $\nu_e$ is 0.3 \cite{vo80}. Thus,
the $\nu_e$ flux is approximately a factor $(0.3/0.35)^\gamma$ times the $\nu_\mu$ flux produced
in the muon decay (i.e. a factor of $\sim 0.7$ for a nucleon spectral index $\gamma=2.7$, again in good
agreement with previous, more detailed calculations \cite{in96}). Hence, 
if the neutrino flavors are redistributed by oscillations, the ($\nu_\mu+
\bar\nu_\mu$) galactic flux will be a factor $(1+0.7\times 2.1/3.1)/3
\simeq 1/2$ of that depicted in figure~4, which neglects the neutrino oscillations.

In order to calculate the galactic neutrino flux induced by the full CR spectrum, the procedure 
follows the same steps already described above for the atmospheric fluxes. The total galactic 
neutrino fluxes are shown in figure 4(a) for $R=1$ and 20 kpc, 
and correspond to the $\Delta\alpha=2/3$ rigidity dependent scenario (with the same parametrization 
used for the previous figures and detailed in section 2). For comparison, the figure also
shows the total atmospheric neutrino flux in the horizontal direction with the prompt
charm component calculated using the GBW model, as well as the Waxman-Bahcall flux predictions for 
the astrophysical diffuse neutrino fluxes produced in their model of GRBs \cite{wa99,ah03}. 
Analogously, figure 4(b) shows the results corresponding to the same total CR spectrum but assuming
it to consist only of protons. In agreement with previous results \cite{in96}, we observe that 
the galactic flux in the direction orthogonal to the plane (corresponding for instance to $R=1$~kpc) 
remains below the atmospheric flux up to the highest energies. Along directions near the galactic plane
(e.g. for $R=20$~kpc), the flux of galactic neutrinos produced in the ISM instead overcomes the 
atmospheric flux at energies larger than $\sim 10^{14}$~eV. The energy at which the galactic
flux actually dominates depends to a large extent on the particular model adopted for charm 
production, as discussed above, and the results naturally also vary according to the assumed ISM 
column density and the zenith angle of arrival direction in the atmosphere. These figures show 
very clearly the significant effect of suppression in the neutrino fluxes that results from 
considering a rigidity dependent scenario. Indeed, above $\sim 10^{15}$~eV the Waxman-Bahcall flux
is found to overcome the background of galactic neutrinos with trajectories nearly contained in the
galactic plane. However, if the same total CR spectrum is taken to be formed by protons alone, the
resulting background flux of galactic neutrinos turns out to be larger than the Waxman-Bahcall flux
up to $\sim 10^{16}$~eV, and then both fluxes stay with comparable magnitude up to the 
highest energies. Hence, rigidity dependent scenarios result in a significantly 
reduced background for the search of astrophysical neutrino sources of 
current interest, such as for instance GRBs or AGNs.         

\section{Conclusions}

We computed in this work the high energy diffuse fluxes of muon 
neutrinos (and antineutrinos) produced by CRs hitting the upper atmosphere or
interacting with the ISM in the Galaxy, showing that the results are 
quite sensitive to both the total CR spectrum and the CR composition assumed. 
Concerning the former, we compared previous results (which adopted an overall
normalization for the CR spectrum actually somewhat below present observational
data, and a single and abrupt slope change representing the knee) with those
obtained from considering a smooth CR spectrum that soundly fits the observations
and reproduces its main relevant features (namely, the knee, the second knee and
the ankle). With respect to the latter, we showed that taking into account a
CR composition that turns heavier above the knee (i.e. in agreement with the scenarios
that explain the knee as due to a rigidity dependent effect) the induced neutrino 
fluxes become significantly suppressed, thus making their detection
harder but also reducing the background for the 
search of other astrophysical neutrino sources. In particular, we compared our
results with the Waxman-Bahcall predictions for the neutrino fluxes produced 
in their model of GRBs, and found indeed that the signal-to-background ratio 
for this model becomes
significantly enhanced when considering a CR composition with the 
contribution of different nuclear species, as opposed to the case in which
the full CR spectrum is assumed to be composed of protons alone. An analogous
effect would lower the neutrino flux produced by the extragalactic component 
(which gives the dominant contribution to the total neutrino flux above 
$E\sim 10^{17}$~eV) if it were not constituted just by protons, as we assumed here.
It is then clear that if the diffuse neutrino fluxes predicted in the GRBs models \cite{wa99}
or the even larger ones associated to some AGN models \cite{st96} do actually exist, 
the observation of atmospheric or galactic CR induced neutrino fluxes above
$10^{14}$~eV will remain probably hopeless.

Let us also mention that it may be of interest to consider 
the atmospheric muon fluxes produced by CRs. For instance, it has
recently been suggested \cite{ge03} that the observation of 
the down-going atmospheric muons with  neutrino telescopes would
provide an indirect measure of the prompt atmospheric neutrino flux,
and hence this could be used to confront the NLO QCD predictions.
Indeed, due to the charmed particle semileptonic decay kinematics,     
it turns out that the prompt muon flux coincides with the prompt neutrino
flux to within $\sim 10\%$, irrespective of the energy and independently of the
model used to treat the atmospheric charm production \cite{ge03}. On the other hand,
the atmospheric conventional muon flux is about a factor of $\sim 5$ larger 
than the conventional muon neutrino flux within the energy 
range of interest of this work, 
and exhibits roughly the same energy dependence \cite{ge03}. Completely analogous 
effects as those described here for the neutrino fluxes should then 
be expected for the atmospheric muons produced by CRs reaching
the earth.          

Finally, we stress the key importance of determining confidently the CR composition 
around and above the knee, since it appears in this context as decisive in 
order to estimate reliably the diffuse high energy neutrino background produced
by CRs. Moreover, solving the CR composition puzzle will also provide a
valuable means of testing the different proposals concerning the origin and
nature of the knee, and will thus shed new light on this long standing problem.

\section*{Acknowledgments}
Work supported by CONICET and Fundaci\'on Antorchas, Argentina. 

\end{document}